\documentclass[prl,aps,epsf,twocolumn,showpacs,10pt,superscriptaddress,groupeadaddress]{revtex4-1}
\usepackage{graphicx,amsfonts,times,bm,amsmath,verbatim,color,array}
\usepackage[colorlinks, allcolors=blue]{hyperref}

\usepackage{natbib}
\usepackage{graphicx,amssymb,amsmath,ifthen,braket,xcolor}
\usepackage{bm}

\begin{document}


\title{Spin-Peierls transition of the dimer phase of the $J_1-J_2$ model: Energy cusp and CuGeO$_3$ thermodynamics}



\author{Sudip Kumar Saha}
\affiliation{S. N. Bose National Centre for Basic Sciences, Block - JD, Sector - III, Salt Lake, Kolkata - 700106, India}

\author{Manoranjan Kumar}
\email{manoranjan.kumar@bose.res.in}
\affiliation{S. N. Bose National Centre for Basic Sciences, Block - JD, Sector - III, Salt Lake, Kolkata - 700106, India}

\author{Zolt\'an G. Soos}
\email{soos@princeton.edu}
\affiliation{Department of Chemistry, Princeton University, Princeton, New Jersey 08544, USA}



\date{\today}

\begin{abstract}

The spin-Peierls transition is modeled in the dimer phase of the spin-$1/2$ chain with exchanges $J_1$, $J_2 = \alpha J_1$ between first 
and second neighbors. The degenerate ground state generates an energy cusp that qualitatively changes the dimerization $\delta(T)$ compared 
to Peierls systems with nondegenerate ground states. The parameters $J_1 = 160$ K, $\alpha = 0.35$ plus a lattice stiffness account for the 
magnetic susceptibility of CuGeO$_3$, its specific heat anomaly, and the $T$ dependence of the lowest gap. 
\end{abstract}

\pacs{}

\maketitle

\textit{Introduction.}\textemdash
The spin-Peierls (SP) transition at $T_{SP} = 14$ K in CuGeO$_3$ crystals discovered by Hase \textit{et al.}~\cite{haseprl1993, *haseprb1993} initiated an intense, 
decade long effort to characterize~\cite{riera1995,fabricius1998,castilla1995,bouzerar1999,nishi1994,vollenkle1967,hidaka1997,lussier1996,martin1996,lorenz1996,liu1995} 
the spin-$1/2$ chains of Cu(II) ions along the crystallographic $c$ axis. As noted in the review by Uchinokura~\cite{uchinokura2002}, 
sizable single crystals made possible detailed structural, thermodynamic and spectroscopic investigations. It soon emerged that the SP transition of CuGeO$_3$ 
is unlike the organic SP systems that Jacobs \textit{et al.}~\cite{jacob1976, *bray1983} modeled successfully as linear Heisenberg antiferromagnets (HAFs) with isotropic
exchange $J_1 > 0$ between 
first neighbors in spin-$1/2$ chains of molecular cation radicals. Moreover, Peierls and SP transitions were broadly compatible with opening a gap in a BCS superconductor. 
CuGeO$_3$ is an outlier whose SP transition is different from previous systems in general, and from organic chains modeled as HAFs in particular.

As pointed out by Riera and Dobry~\cite{riera1995}, an improved fit of the CuGeO$_3$ magnetic susceptibility $\chi(T)$ is obtained by adding an exchange $J_2 = \alpha J_1$ 
between second neighbors. The $J_1-J_2$ model has its own literature. The frustration parameter $\alpha= 0.35$ and $J_1 = 160$ K that fits the high-$T$ 
susceptibility~\cite{fabricius1998} is in the dimer phase that runs from the quantum critical point~\cite{nomura1992} $\alpha_c = 0.2411$ where a singlet-triplet 
gap $\Delta (\alpha)$ opens to the Majumdar-Gosh (MG) point~\cite{ckm69b} $\alpha = 0.50$ where the exact ground state is known. But $\alpha < \alpha_c$ has been 
advocated for other reasons~\cite{castilla1995,bouzerar1999} and while frustration is clearly important, the magnitude of $\alpha$ has not been definitively determined~\cite{uchinokura2002}. 
Inelastic neutron scattering~\cite{nishi1994} indicates an exchange $J_\perp \sim 0.1J_1$ between spins in adjacent chains; a 1D model may not be appropriate for CuGeO$_3$. 
Another complication is that single crystals, $T_{SP}$ and properties depend somewhat on growth conditions. Most analysis is in terms of the $300$ K structure 
reported by Vollenkle \textit{et al.}~\cite{vollenkle1967} with space group $Pbmm (D_{2h} ^5)$. The refined structure reported by Hidaka \textit{et al.}~\cite{hidaka1997} 
has lower symmetry and larger unit cell. 

We discuss in this paper the SP transition in the dimer phase of the $J_1-J_2$ model and show that the dimer phase is the key difference from previous 
Peierls or SP systems. The ground state of the rigid lattice is doubly degenerate with spontaneously broken inversion symmetry, while the ground state of other 
systems is nondegenerate. That includes the inorganic and organic systems dating back to Krogmann salts that Pouget has recently reviewed~\cite{pouget2016,*pouget2017}. 

We consider the MG point where the degeneracy occurs in finite chains and $\alpha = 0.35$ where the degeneracy is in the thermodynamic limit. Degeneracy 
has not been discussed in SP modeling of finite chains, which indeed are nondegenerate. Although $\alpha = 0.35$ is motivated by CuGeO$_3$ and we make 
contact below with its thermodynamics, our principal goal is the SP transition of the dimer phase of the $J_1-J_2$ model.  The basic model has only 
three parameters: $J_1$ and $\alpha$ specify the spin chain. Dimerization is opposed by a harmonic lattice with linear spin-phonon coupling. A realistic 
model of CuGeO$_3$ is clearly more complicated.\\ 

%
\textit{Energy cusp and equilibrium.}\textemdash
We take $J_1$ as the unit of energy, frustration $\alpha= J_2/J_1$ and dimerization $\delta$. The dimerized $J_1-J_2$ model for spin $S_r = 1/2$ at site $r$ 
and periodic boundary conditions is
\begin{equation}
	H(\alpha,\delta) = \sum_{r} \left( 1+\delta (-1)^r  \right)   \vec{S}_r \cdot \vec{S}_{r+1} + \alpha \sum_{r} \vec{S}_{r} \cdot \vec{S}_{r+2}.
\label{eq:j1j2}
\end{equation}
The ground state of $H(\alpha,0)$ is doubly degenerate in the dimer phase. Finite $\delta$ breaks inversion symmetry at sites and increases the 
singlet-triplet gap $\Delta(\alpha, \delta)$ but does not change the chain length in systems with periodic boundary conditions.

In reduced units, the free energy per spin in the thermodynamic limit is
\begin{equation}
	A(T,\alpha,\delta) = -T \ln{Q(T,\alpha,\delta)}.
\label{eq:free_energy}
\end{equation}
The corresponding expression for finite systems has $N^{-1} \ln{Q(T,\alpha,\delta,N)}$. The Boltzmann sum in the partition function is over the 
eigenstates $E_j (\alpha,\delta,N)$ of the $N$-spin system. The adiabatic approximation for a lattice with potential energy per site $\delta^2/2\varepsilon_d$ 
is standard practice. The equilibrium condition that minimizes the total free energy is
\begin{equation}
	\frac {\delta(T,\alpha)} {\varepsilon_d} = -\left( \frac {\partial A(T,\alpha,\delta)} {\partial \delta}  \right)_{\delta(T,\alpha)}.
\label{eq:equilibrium_condition}
\end{equation}
This relation holds for finite $N$ as well as for other Peierls systems. Dimerization $\delta(T,\alpha)$ is largest at $T = 0$ and decreases to zero at $T = T_{SP}$.

The ground state energy per site, $E_0(\alpha,\delta,N)$, is the free energy at $T = 0$. The two VB diagrams with $N/2$ singlet pairs between neighbors are the $\delta= 0$ 
ground states at the MG point.  The exact $\delta= 0$ energy and slope are both $-3/8$. The $\alpha= 0.35$, $\delta= 0$ ground state has avoided crossings at finite $N$. 
The cusp in the thermodynamic limit is the amplitude $B(\alpha)$ of the bond order wave~\cite{mkumarbndord2010},
\begin{equation}
\begin{aligned}
   &	-\left( \frac {\partial E_0 (\alpha,\delta)} {\partial \delta} \right)_0=B(\alpha) =\\ 
   & 	\langle \psi_1(\alpha) \vert \sum_{r} (-1)^r \vec{S}_r \cdot \vec{S}_{r+1}
	\vert \psi_{-1}(\alpha) \rangle /N.
\label{eq:bond_order}
\end{aligned}
\end{equation}
The degenerate ground states $\psi_{\pm 1} (\alpha)$  are even and odd, respectively, under inversion at sites. We obtain smaller $B(0.35) = 0.0783$ compared to $B(0.5) = 0.375$. 
$B(\alpha)$ is exponentially small near $\alpha_c = 0.2411$ and is intrinsic to the dimer phase.

\begin{figure}
\includegraphics[width=\columnwidth]{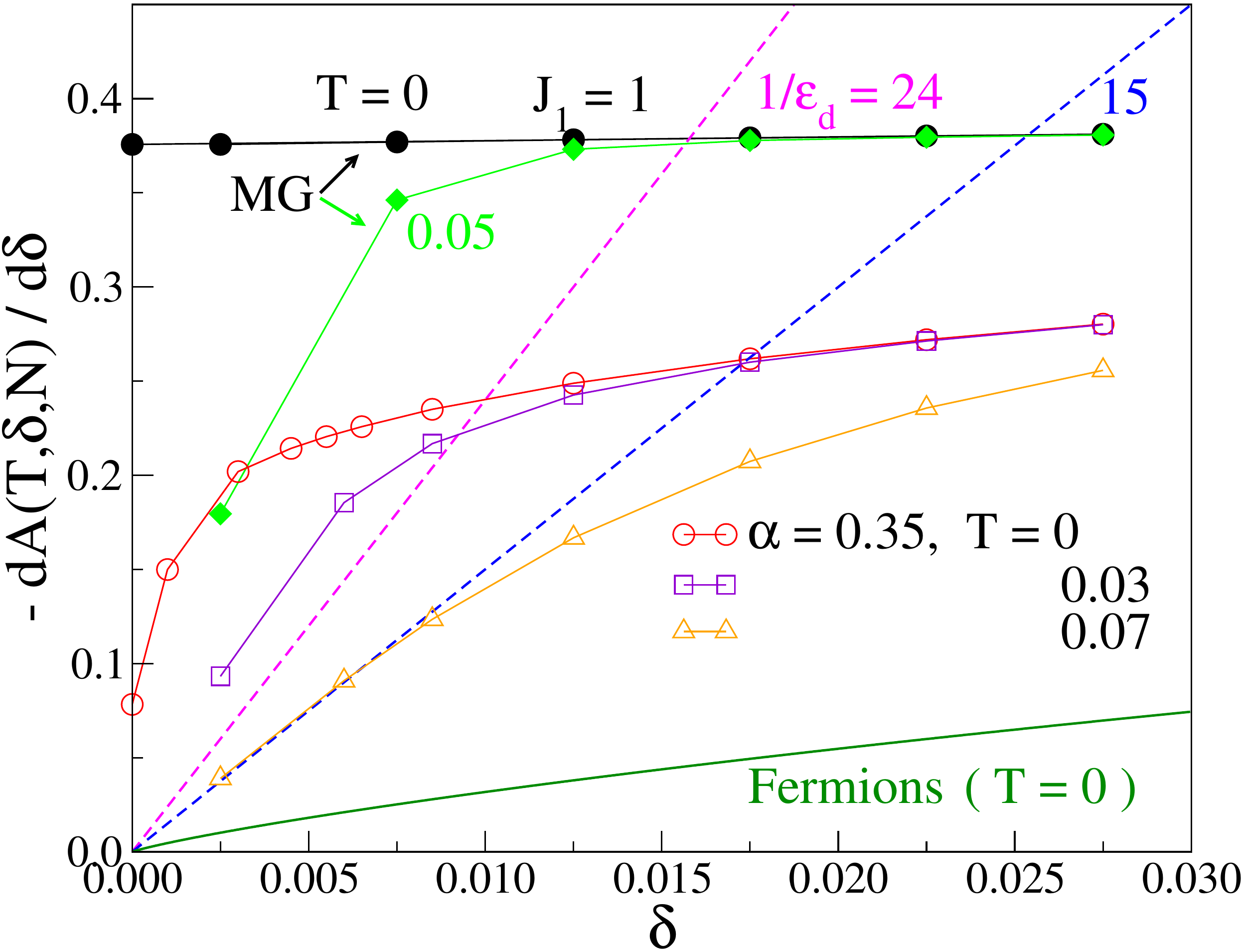}
\caption{\label{fig1} Free-energy derivatives vs. dimerization in systems of $32$ spins, with cusp $B(\alpha)$ at $\delta= T = 0$. The crossing points with the lines 
$\delta/\varepsilon_d$ satisfy the equilibrium condition, Eq.~\ref{eq:equilibrium_condition}, at $\delta(\alpha,T)$. The points for $\alpha= 0.35$, $T = 0$, $\delta< 0.015$ are for $N = 96$.}
\end{figure}
Fig.~\ref{fig1} shows $-(\partial A/\partial \delta)$ as functions of $\delta$ at $\alpha= 0.50$ and $0.35$. The thermodynamic limit is easily reached at $\alpha= 0.50$ but 
requires $N \sim 100$ at $\alpha= 0.35$, $\delta= 0$. The dashed $\delta/\varepsilon_d$ lines are shown for $1/\varepsilon_d = 15$ and $24$, which leads respectively 
to $\delta(0) = 0.025$ and $0.016$ at $\alpha= 0.50$. The $\alpha= 0.35$ lines intersect at smaller $\delta(0)$. By contrast, the right-hand side of Eq.~\ref{eq:equilibrium_condition} 
is proportional to $\delta$ in Peierls systems with a nondegenerate ground state. For example, the ground state of a half-filled band of noninteracting spinless fermions leads to
\begin{equation}
	-\left(\frac {\partial E_0(\delta)} {\partial \delta} \right)=\frac {2\delta \left[K(q)-E(q) \right]} {\pi q^2}.
\label{eq:spinless}
\end{equation}
Here $q^2 = 1 - \delta^2$ and $K(q)$, $E(q)$ are complete elliptic integrals of the first and second kind. The fermion result in Fig.~\ref{fig1} is qualitatively different 
since it passes through the origin. Nevertheless, finite $\delta(0)$ is guaranteed for finite $\varepsilon_d$ after dividing both sides of Eq.~\ref{eq:equilibrium_condition} 
by $\delta$ because $K(q)$ has a logarithmic divergence at $\delta= 0$. The half-filled tight binding (H\"uckel) band has $4\delta$ instead of $2\delta$ in Eq.~\ref{eq:spinless}. 
Previous Peierls or SP systems~\cite{jacob1976, *bray1983, pouget2016, *pouget2017} have nondegenerate ground states; 
the $E_0(\delta)$ expansion has only even powers of $\delta$. \\

%
\textit{SP transition.}\textemdash
Dimerization at finite $T$ follows from the equilibrium condition. The $T = 0.03$ or $0.05$ lines in Fig.~\ref{fig1} intersect $\delta/\varepsilon_d$ at smaller $\delta(T)$, while 
the $T = 0.07$ line for $\alpha = 0.35$ passes through the origin. We obtain the thermodynamics of $H(\alpha,\delta)$ by exact diagonalization (ED) to $N = 24$ for $\delta = 0$ and 
$N = 20$ for $\delta> 0$. We use the density matrix renormalization group (DMRG) for the ground state properties of larger systems whose thermodynamics is found by a combination of 
ED and DMRG~\cite{sudip19}. The energy spectrum ${E(\delta,N)}$ is truncated at a cutoff $W_C(\delta,N)$ based on $S(T,\delta,N)/T$, where $S$ is the entropy per site. $S(T,\delta,N)$ 
is a lower bound on the thermodynamic limit, $S(T,\delta)$, due to finite size gaps at low $T$ and truncation at high $T$. The maximum of $S(T,\delta,N)/T$ at $T^\prime(\delta,N)$ 
is the best approximation of $S(T^\prime,\delta)$ for the truncated spectrum. Accordingly, we keep the lowest few thousand states needed to converge, or almost to converge, the maximum of $S(T,\delta,N)/T$.
\begin{figure}
\includegraphics[width=\columnwidth]{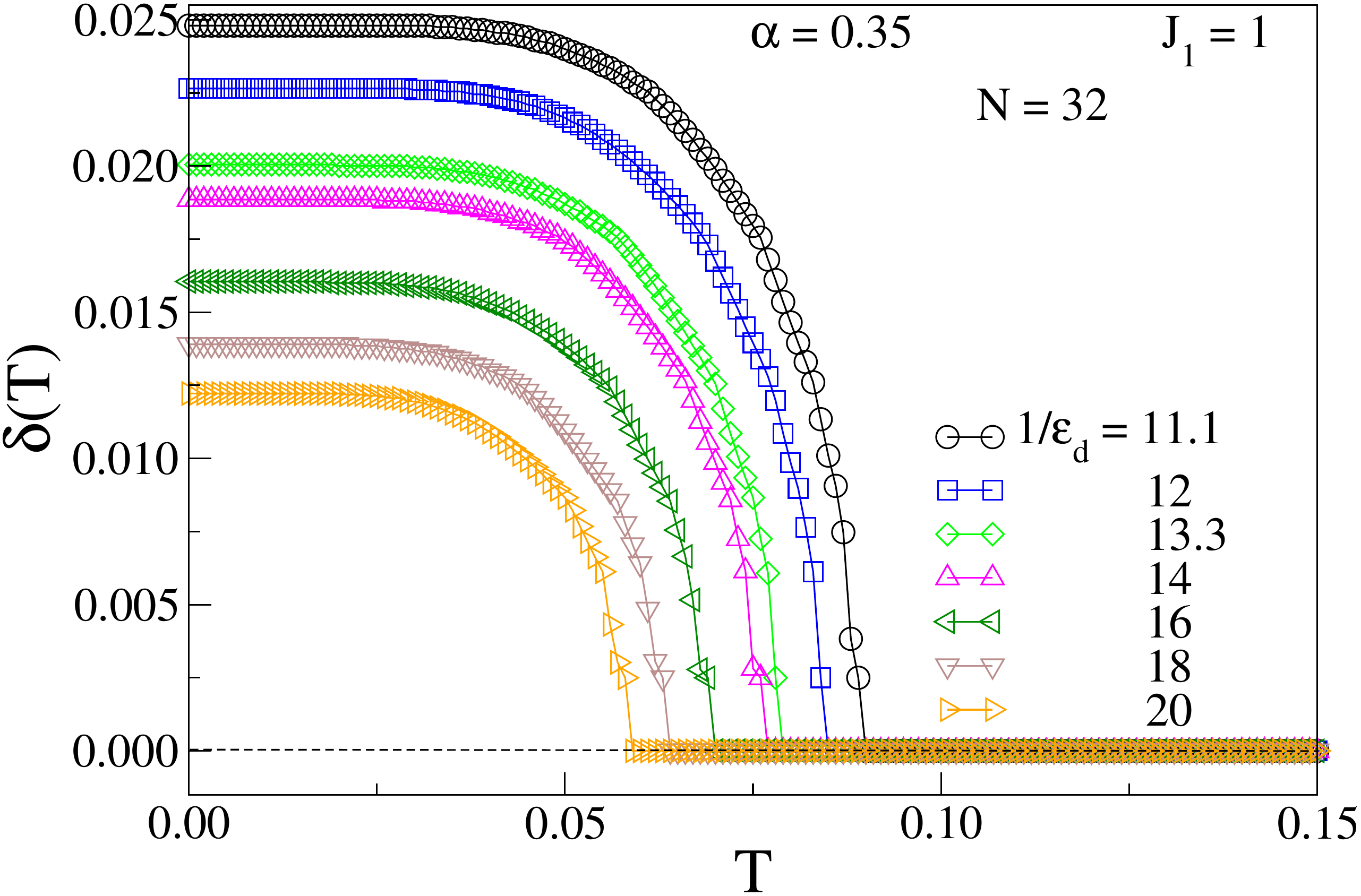}
\caption{\label{fig2} 
	Equilibrium dimerization $\delta(T)$ and SP transition $T_{SP}$ of $32$-spin systems with $\alpha= 0.35$ and the indicated stiffness $1/\varepsilon_d$.}
\end{figure}

Fig.~\ref{fig2} shows $\delta(T)$ vs $T$ at $\alpha= 0.35$ and $N = 32$. Each point is an intersection as in Fig.~\ref{fig1}. Comparison with results for $N = 20$ and $24$, 
as well as fewer points at $N = 48$ and $64$, indicate that $N = 32$ is close to the thermodynamic limit. As expected, $\delta(0)$ and $T_{SP}$ decrease with increasing 
stiffness $1/\varepsilon_d$. Similar $\delta(T)$ vs. $T$ curves with higher $T_{SP}$ are found at $\alpha= 0.50$.
\begin{figure}
\includegraphics[width=\columnwidth]{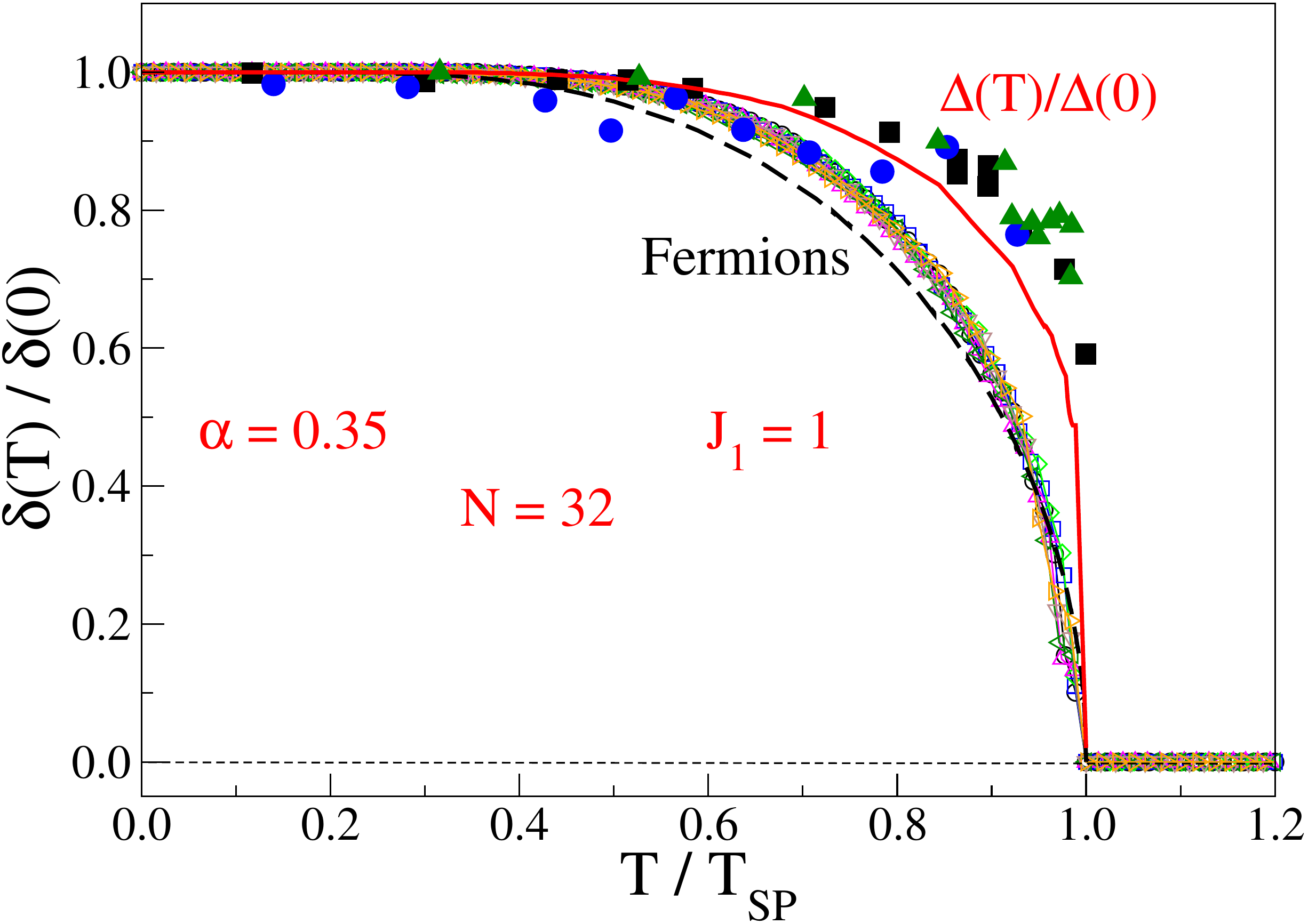}
\caption{\label{fig3}
Scaled $\delta(T)/\delta(0)$ vs. $T/T_{SP}$ curves from Fig.~\ref{fig2}. The solid line is the singlet-triplet gap ratio, $\Delta(\alpha,\delta(T))/\Delta(\alpha,\delta(0))$. 
	The symbols are inelastic neutron data: blue filled circles (Ref.~\onlinecite{nishi1994}), red filled squares (Ref.~\onlinecite{lussier1996}) and green filled triangles
	(Ref.~\onlinecite{martin1996}). The dashed line is the gap ratio $\delta(T)/\delta(0)$ for a half-filled band of noninteracting fermions 
	with small $\delta(0)$.}
\end{figure}

The choice $1/\varepsilon_d = 11.1$ returns $T_{SP} = 0.09 = 14.4$ K, in the observed range for CuGeO$_3$ crystals. The calculated dimerization is $\delta(0) = 0.025$. 
Fig.~\ref{fig3} shows that $\delta(T)/\delta(0)$ vs. $T/T_{SP}$ falls on a single line for all curves in Fig.~\ref{fig2}. There is little dependence on $\varepsilon_d$ in stiff 
lattices with small $\delta(0)$. The ratio of the gaps $\Delta(\alpha,\delta(T))$ at $T$ and $0$ in Fig.~\ref{fig3} is evaluated at $\delta(T)$ along the $1/\varepsilon_d = 11.1$ 
line. The gap ratio, not shown, for $\Delta(\alpha)$ at $T \geq T_{SP}$ is less than $3\%$ in the thermodynamic limit.

We include in Fig.~\ref{fig3} gap ratios from inelastic neutron scattering~\cite{nishi1994,lussier1996,martin1996}. The dashed line is $\delta(T)/\delta(0)$ for a half-filled band of 
noninteracting fermions, spinless or spin-1/2, with small $\delta(0)$ and, more 
importantly in the present context, gap $4\delta(T)$. The SP transition of the HAF, approximations and connections to BCS are discussed in Ref.~\onlinecite{jacob1976, *bray1983}. The neutron data disagreed with general expectations. The scaled gaps are instead consistent with an SP transition in the dimer phase.

The data in Fig.~\ref{fig3} are specific to CuGeO$_3$. We turn to thermodynamics for two reasons. One is to seek additional evidence for the characteristic $\delta(T)$ of 
the dimer phase. The other is to propose that the dimer phase of the $J_1-J_2$ model is the 
proper starting point for CuGeO$_3$ modeling. \\

\begin{figure}
\includegraphics[width=\columnwidth]{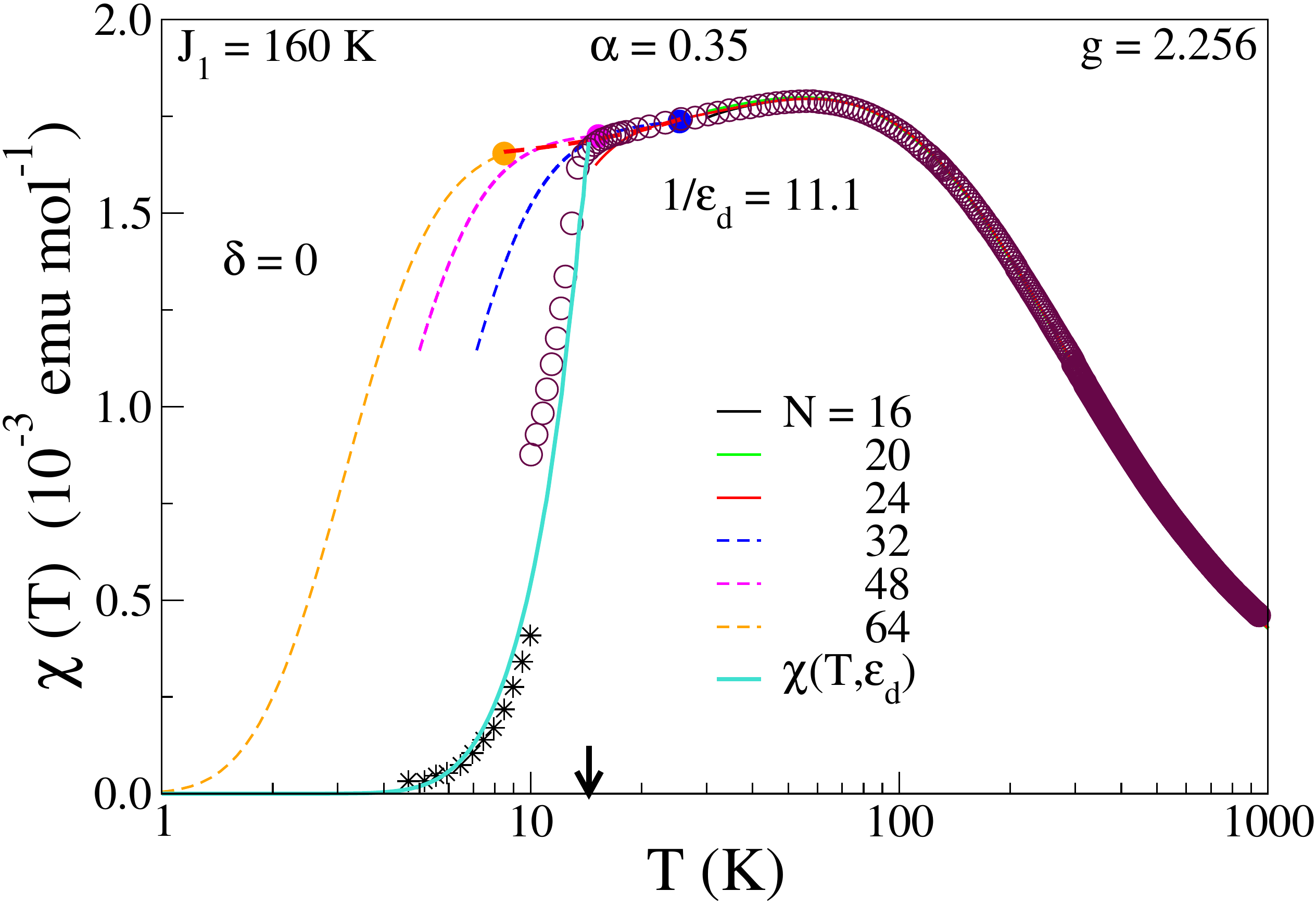}
\caption{\label{fig4}
Molar spin susceptibility $\chi(T)$ of CuGeO$_3$. The solid $\delta= 0$ line is ED; dashed lines are DMRG up to $T^\prime(N)$ discussed in the text. 
	The solid line up to $T_{SP}$ is $\chi(T,\varepsilon_d)$ evaluated at the equilibrium $\delta(T)$. The $T \geq  10$ K data to $950$ K is from 
	Ref.~\onlinecite{fabricius1998}; the $T \leq 10$ K data from Ref.~\onlinecite{haseprl1993, *haseprb1993}.}
\end{figure}

%
\textit{Susceptibility and specific heat.}\textemdash
The $\chi(T)$ data~\cite{fabricius1998} in Fig.~\ref{fig4} from $T = 10$ to $950$ K was kindly provided in digital form by Professor Lorenz.  ED for $N = 18$ reaches the thermodynamic 
limit at $T < T_{max} = 56$ K, the maximum that was used to fix the parameters~\cite{fabricius1998} $J_1 = 160$ K, $\alpha= 0.35$ and (from ESR)
$g = 2.256$. ED to $N = 24$ reaches the 
limit by $T \sim 25$ K. DMRG results are shown as dashed lines that terminate at the solid point $T^\prime (N)$ that are the best approximation to the thermodynamic limit for 
the truncated energy spectrum. The limit for $N = 48$ holds for $T > 10$ K, and $\chi(T,0)$ closely matches the $T > T_{SP}$ data. Bouzerar \textit{et al.}~\cite{bouzerar1999} discuss 
reasons for choosing $\alpha$ slightly less than $\alpha_c$ in Eq.~\ref{eq:j1j2}, which leads~\cite{castilla1995} to $T_m > 100$ K. They address 
the issue by treating the observed~\cite{nishi1994} 
interchain exchange $J_\perp \sim 0.1J_1$ at a mean field level; even so their $\chi(T)$ fit (Fig.1, ref.~\onlinecite{bouzerar1999}) is rather approximate below $100$ K.

The calculated $\chi(T,\varepsilon_d)$ at $T < T_{SP} = 14.4$ K is evaluated with
$\delta(T)$ along the $1/\varepsilon_d = 11.1$ line in Fig.~\ref{fig2}. The 
thermodynamic limit is 
easily reached for dimerized systems, here with $\delta(0) = 0.025$. Data~\cite{haseprl1993, *haseprb1993} up to $10$ K are shown as crosses. The $10$ K points do not match. The $T < T_{SP}$ 
fit is for the basic 1D model, Eq.~\ref{eq:j1j2}, with isotropic exchange.

\begin{figure}
\includegraphics[width=\columnwidth]{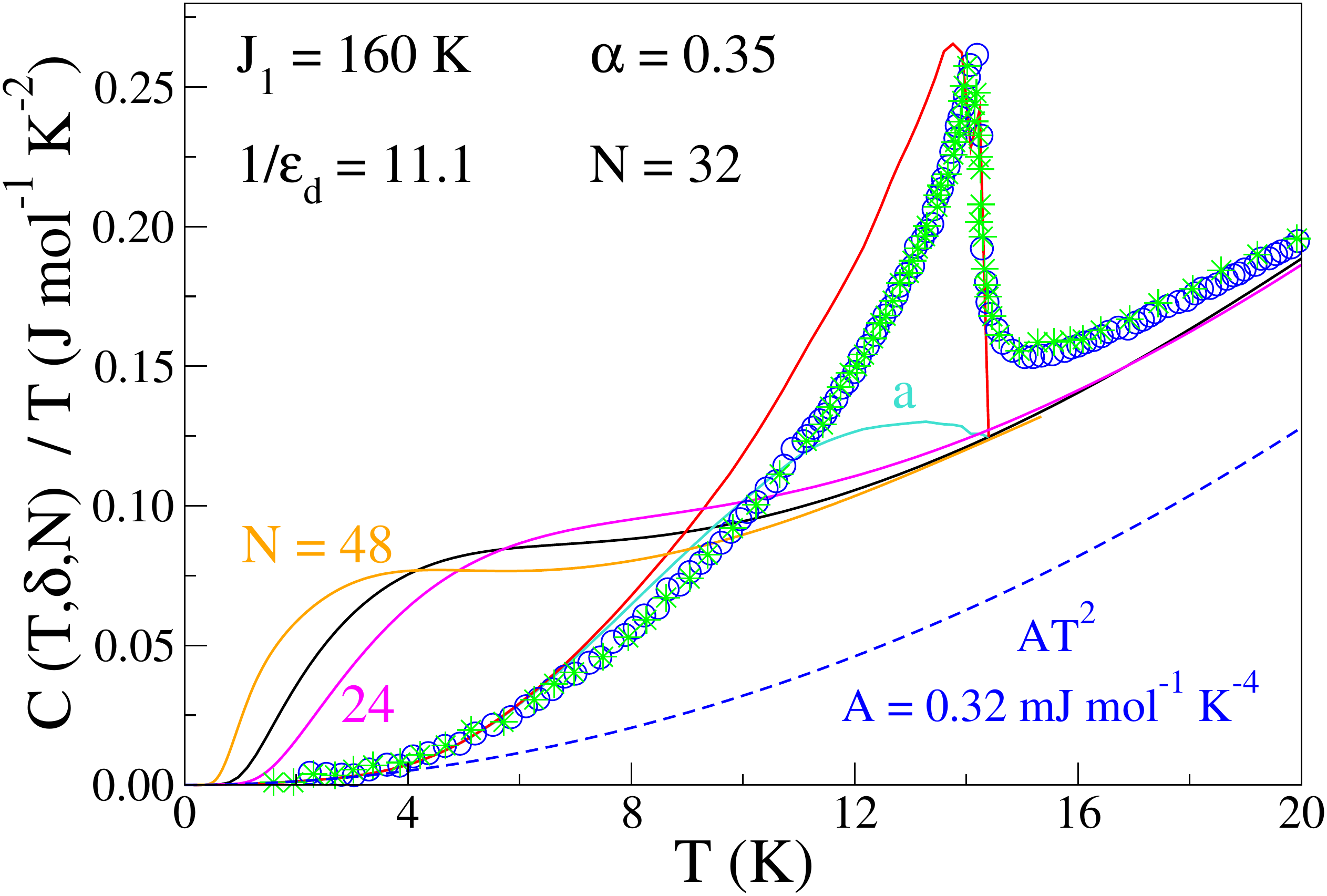}
\caption{\label{fig5}
	Molar specific heat C(T) of CuGeO$_3$ shown as the entropy derivative $S^\prime = C/T$: blue circles from Ref.~\onlinecite{lorenz1996}, green stars and A 
	from Ref.~\onlinecite{liu1995}. The calculated $C(T,0)/T$ curves are ED for $N = 24$, DMRG for $N = 32$ and $48$. The equilibrium $C(T,\varepsilon_d)/T$ is 
	Eq.~\ref{eq:specific_heat} 
	whose first term is labeled (a); the second term with the distinctive $\partial \delta(T)/\partial(T)$ of the dimer phase describes the anomaly.}
\end{figure}

The measured~\cite{lorenz1996, liu1995} specific heat $C(T)$ to $20$ K is shown in Fig.~\ref{fig5} as the entropy derivative $S^\prime = C/T$. The dashed line is 
the lattice (Debye) contribution, $AT^2$, with~\cite{liu1995} $A= 0.32$ mJ mole$^{-1}$K$^{-4}$. The anomaly is sharper and better resolved than in organic SP 
systems~\cite{wei1977}. The equilibrium $C(T,\varepsilon_d)$ has two contributions below $T_{SP}$,
\begin{equation}
\begin{aligned}
	& C(T,\varepsilon_d)=C(T,\delta(T))+ \\
	& \qquad \frac {\partial \delta} {\partial T} \left[ \left( \frac {\partial E(T,\delta)} {\partial \delta} \right)_T +\frac {\delta(T)} {\varepsilon_d} \right].
\label{eq:specific_heat}
\end{aligned}
\end{equation}
The first term is evaluated at $\delta(T)$ along the $1/\varepsilon_d = 11.1$ line in Fig.~\ref{fig2}. The second term tracks $\delta(T)$ variations. 
The calculated $C(T,\varepsilon_d)/T$ and $C(T,0)/T$ are shown in Fig.~\ref{fig5}. The low-$T$ behavior at $\delta= 0$ is a finite size effect. Since 
the gap $\Delta(\alpha,N) > \Delta(\alpha)$ initially decreases $C(T,\alpha,N)/T$, entropy conservation requires increased $C(T,\alpha,N)/T$ before 
converging from above to the thermodynamic limit. The $N = 48$ and $24$ gaps are smaller and larger, respectively, than $N = 32$. The thermodynamic limit 
is almost reached by $12$ K for $N = 32$.

The $C(T,\delta(T))/T$ term generates the curve labeled (a) in Fig.~\ref{fig5}. The $\partial \delta(T)/\partial T$ term is mainly responsible for the sharp 
anomaly. It was computed by fitting $\delta(T)$ in Fig.~\ref{fig3} to a smooth curve and taking the derivative. The area $C(T,\varepsilon_d)/T$ to $T_{SP}$ is 
within $5\%$ of the accurately known area for $\delta= 0$. The anomaly is directly related to $\delta(T)$ of the dimer phase. The adiabatic and mean field 
approximations for the lattice enforce $\delta= 0$ for $T > T_{SP}$; this 
oversimplification is a general problem for modeling any transition and
has long been recognized. The measured area 
(or entropy) at $20$ K is within a few percent of the $\delta= 0$ entropy. Lattice fluctuations above $T_{SP}$ imply reduced entropy below $T_{SP}$. 
Overall, the anomaly is fit almost quantitatively.  \\

%
%
\textit{Conclusions.}\textemdash
We have analyzed the SP transition of the dimer phase of the $J_1-J_2$ model. Dimerization $\delta> 0$ lifts the degeneracy and generates a cusp $B(\alpha)$ 
at $\delta= 0$ in the ground-state energy $E_0(\delta)$. The cusp differentiates the dimer phase from previously studied Peierls and SP models 
with nondegenerate ground states. Such models have $E_0^\prime(0) = 0$ and transitions due to divergent curvatures $E_0^{\prime \prime}(0)$. 
Their transitions illustrate shared features among otherwise quite dissimilar quasi-1D systems.  

The equilibrium condition, Eq.~\ref{eq:equilibrium_condition}, for dimerization returns a different $\delta(T)$ when $E_0(\delta)$ has a cusp. 
The temperature dependence of $\delta(T)/\delta(0)$ or of the scaled gap in Fig.~\ref{fig3} is characteristic of the dimer phase. The specific heat 
anomaly in Fig.~\ref{fig5} is directly related to $\partial \delta(T)/\partial T$. The parameters $J_1 = 160$ K and $\alpha= 0.35$ in Eq.~\ref{eq:j1j2} 
plus the stiffness $1/\varepsilon_d = 11.1$ in Eq.~\ref{eq:equilibrium_condition} account for several features of the SP transition of CuGeO$_3$. We 
are applying the basic model to other data and think that the dimer phase is the starting point for detailed modeling these quasi-1D spin-$1/2$ chains.
\\

\begin{acknowledgments}
We thank T. Lorenz for providing us the $\chi(T)$ data. 
ZGS thanks D. Huse for several clarifying discussions.
MK thanks DST India for financial support through a Ramanujan fellowship.
SKS thanks DST-INSPIRE for financial support.
\end{acknowledgments}

\bibliography{ref_thermo2}
\end{document}